\documentclass{emulateapj}
\slugcomment{Accepted for publication in ApJ}
\lefthead{Laha et al.}

\usepackage{color}

\usepackage{rotating}
\usepackage{lscape}

\def\unit #1{\,{\rm #1}}

\newcommand\kms{\rm \,\unit{km\,s^{-1}}}

\newcommand\cmsqi{\rm \,\unit{cm^{-2}}}

\newcommand\ksec{\rm \,\unit{ks}}

\newcommand\cmcubei{\rm \,\unit{cm^{-3}}}

\newcommand\kev{\rm \,\unit{keV}}

\newcommand\funit{\rm \,erg\,cm^{-2}\,s^{-1}}

\newcommand\lunit{\rm \,erg \,s^{-1}}

\newcommand\lunita{\rm\,erg \,s^{-1}\AA^{-1}}

\newcommand\punit{\rm \,photons\,cm^{-2}\,s^{-1}}

\newcommand\xiunit{\rm \,erg\,cm\,s^{-1}}

\newcommand\ledd{L_{\rm \, Edd}}

\newcommand\lambdaedd{\lambda_{\rm \, Edd}}

\newcommand\lbol{L_{\rm \, bol}}

\newcommand\msol{M_{\odot}}

\newcommand\mbh{M_{\rm BH}}

\newcommand\msolyi{M_{\odot}\,\rm yr^{-1}}

\newcommand\nh{N_{\rm H}}

\newcommand\nhwa{N^{\rm WA}_{\rm H}}

\newcommand\nhwe{N^{\rm WE}_{\rm H}}

\newcommand\ev{\unit{\, eV}}

\def\ten#1#2{#1{\times10^{#2}}}

\def\onlyten#1{10^{#1}}

\newcommand\asca{{\it ASCA}}

\newcommand\chandra{{\it Chandra}}
\newcommand\rosat{{\it ROSAT}}

\newcommand\swift{{\it Swift}}
\newcommand\xmm{{\it XMM-Newton}}

\usepackage{graphicx}

\begin{document}

\title{ X-ray warm absorption and emission in the polar scattered Seyfert 1 galaxy Mrk~704}

\author{Sibasish Laha\altaffilmark{1}, Gulab C. Dewangan\altaffilmark{1}, Ajit K. Kembhavi\altaffilmark{1}}\altaffiltext{1}{{Inter University Centre for Astronomy and Astrophysics}; {\tt email: laha@iucaa.ernet.in ; gulabd@iucaa.ernet.in} }

\begin{abstract}
  We present a detailed study of the ionised environment of the
  Seyfert 1 galaxy Mrk~704 using medium and high resolution X-ray
  spectra obtained with a long \xmm{} observation. { The $0.3-10\kev$   continuum,
  well described by a power-law ($\Gamma \approx 1.86$) and two
  blackbodies ($\rm kT\approx0.085$ and $0.22\kev$), is found to be affected by a neutral partial covering absorption ($N_H \approx 10^{23} \cmsqi$, covering fraction  $\approx 0.22$) and two warm absorber components.  }
%We find clear
%  signatures of two components of warm absorbing medium outflowing at
%  different velocities and having different ionisation parameters and
%  column densities. 
  We identify a low ionisation, $\xi \sim 20 \xiunit$, and high
  outflow velocity, $v \sim 1350 \kms$, phase producing the \ion{O}{6}
  and Fe M-shell unresolved-transition array (UTA). An additional high
  ionisation warm absorbing phase with $\xi\sim 500 \xiunit $ and low
  outflow velocity, $v\sim 540 \kms$, gives rise to absorption
  features due to \ion{O}{7}, \ion{O}{8}, \ion{N}{6}, \ion{N}{7} and
  \ion{C}{6}. We also detected %signatures of
 weak emission lines of
  He-like triplets from \ion{O}{7} and \ion{N}{6} ions, thus making
  Mrk~704 a Seyfert 1 galaxy with both warm absorption and 
  emission. The emission lines are well described by two warm
  emitting, photoionised media with different densities but comparable
  $\xi$, suggesting discrete clouds of warm emission. The high density
  phase ($n_e \sim 10^{13}{\cmcubei}$) responsible for the resonance
  lines appears to outflow at high velocity $\sim 5000 \kms$. The low
  velocity, low density phase is likely similar to the X-ray line
  emitting regions found in Seyfert 2 galaxies. The physical
  conditions of warm emitters and warm absorbers suggest that these
  clouds are similar but observed in absorption along our line of
  sight and in emission at other lines of sight. { The
  unique line of sight passing close to the torus opening angle is
  likely responsible for the neutral partial covering absorption and
  our view of emission lines due to the suppressed continuum in this
  polar scattered Seyfert 1 galaxy.}

\end{abstract}

\vspace{0.5cm}


\section{INTRODUCTION}

It is well known that the $0.1-10\kev$ spectra of many Active Galactic
Nuclei (AGN) are modified by partially ionised material along our line
of sight and intrinsic to the source. Such X-ray absorbing clouds
which significantly affect the observed X-ray spectrum have been named
the `` partially ionised absorber'' or the ``warm absorber"
\citep{1984ApJ...281...90H}.  The availability of high resolution
grating X-ray spectra with \xmm{} and \chandra{} have greatly improved
our understanding of discrete absorption and emission features from
AGN.

{Earlier studies on Seyfert galaxies by
\cite{2005A&A...431..111B} and \cite{2005A&A...432...15P} revealed
that around half of the type 1 objects show signatures of warm
absorption in their X-ray spectra}. These warm absorbers give rise to
narrow absorption lines and edges, from elements at a wide range of
ionisation stages (see e.g., \cite{2000A&A...354L..83K,
2000ApJ...535L..17K,2005A&A...431..111B}).  The most prominent of the
absorption features are from the H-like and He-like ions of C, N, O,
Ne and lower ionisation states of Fe. For example \ion{O}{7}
($0.74\kev$), \ion{O}{8} ($0.87\kev$), \ion{Ne}{10} ($1.36\kev$) and
Fe (UTA at $\sim 0.7 \kev $) are primary absorbers above $0.5\kev$
while \ion{C}{5} ($0.39\kev$) and \ion{C}{6} ($0.49\kev$) dominate for
energies below it \citep{1984ApJ...281...90H, 1994MNRAS.268..405N,
1997ASPC..128..173R, 1998ApJS..114...73G}. These lines and edges are
sensitive diagnostics of the ionisation structure and kinematics of
the gas.

The measured blueshift of the absorption lines with respect to the
systemic velocity imply that these absorbers are outflowing with
moderate velocities in the range of $\sim 100-1000 \kms$. {
In some AGNs, high velocity outflows have also been detected \citep[e.g.,][]{2003ApJ...593L..65R,2004A&A...413..921D,2005ApJ...618L..87D,2007ApJ...670..978B}.}
%Chartas et al. 2002, Reeves et al. 2003, Dadina \& Cappi 2004,
%Dasgupta et al. 2005, Braito et al. 2007).  
The resulting mass outflow
rate can be a substantial fraction of the accretion rate required to
power the AGN. Thus, warm absorbers can be dynamically important and
the knowledge of their state, location, geometry and dynamics would
help in understanding the central engines of AGN
\citep{1995AAS...186.4501M}.

We expect the warm absorbers also to emit
in soft X-rays \citep{1993ApJ...411..594N}. However, most Seyfert 1s do
not show clear evidence for such warm emission lines. This may be due to our
direct view of the central power source. According to the {Unified
Model} of AGN, Seyfert 1s are those in which the continuum emitted
from the central engine is viewed directly. This continuum flux being
orders of magnitude greater than the emission line fluxes the latter
is not detectable even though it may be present. On the other hand,
the primary continuum from Seyfert 2s is obscured by an optically
thick torus and the narrow emission lines in the soft X-rays are
detectable on top of an attenuated weak X-ray continuum.  In their
detailed study of the emission line spectrum of the Seyfert 2 galaxy NGC
1068, \cite{2002astro.ph..3021K} have shown that the emission lines
arise from clouds with a large range in ionisation parameters and
column densities, typical of warm absorbers. This suggests that the
emission line spectrum of NGC~1068 arises from clouds that would be observed as warm absorbers if the source were a Seyfert 1 galaxy.

The orientation of some AGNs is intermediate between Seyfert 1 and
2. That is, they are viewed with an inclination comparable with the
``torus'' opening angle, hence the line of sight to the nucleus passes
through the upper layers of the torus. The emission from the central
engine therefore suffers only a moderate extinction through the torus
rim \citep{2002MNRAS.335..773S}, and their polarised spectra are
dominated by polar scattering. Such an AGN is called a polar scattered
Seyfert 1 (PSS) galaxy. 
%{\color {red}In a source like Mrk~704, with an
%intense cold absorption revealed by the ASCA and Swift data
%\cite{2008ApJ...673...96A} these emission features could be more
%evident, only in the case the cold absorber is more intern than the
%emitter.}
 In their detailed study of the optical polarisation spectra
of Seyfert 1 nuclei, Smith et al. have found a sample of such
sources. They have shown that such AGN exhibit polarised broad
emission lines, characteristic of a Seyfert 2 as well as Seyfert 1
spectra in total light. These objects are of particular interest in
the context of the unification scheme since they define a critical
viewing angle at which both Seyfert 1 and Seyfert 2 characteristics
are manifested in the spectrum. Due to their unique viewing angles,
PSS galaxies are expected to show interesting absorption and emission
features. It is this unique viewing angle that led us to perform a
detailed study of the nuclear environment of a PSS galaxy.

Mrk~704 is a
PSS galaxy {\citep{2004MNRAS.350..140S}} and is located at $z=0.0292$. It is a high
Galactic latitude bright X-ray source, detected with \rosat{}
\citep{2000AN....321....1S}. It has also been detected with
\asca{}, \swift{} XRT and BAT. \cite{2008ApJ...673...96A} have modeled
the \asca{} and \swift{} XRT \& BAT spectral data with a partial
covering absorption model (covering fraction $\sim 0.5$ and a high
column density $\nh \sim \ten{1.5}{23}\cmsqi $) modifying a
power-law with $\Gamma \sim 1.36$. They also detected an iron line
with an equivalent width of $160\ev$.  As a part of systematic study
of a sample \cite{2008RMxAC..32..131J} have studied the RGS spectra of
Mrk~704 obtained from a short ($\sim 25\ksec$) \xmm{} observation and
found two distinct components of warm absorbers. Here, we perform a
detailed X-ray spectral study of Mrk~704, both medium and high resolution, based on a long $\sim 100\ksec$ \xmm{}
observation.

 The paper is organised as follows. We describe the observation and
data reduction in Section 2 and spectral analysis and modeling in Section
3. We discuss our results in Section 4, followed by a summary in
Section 5. {Throughout this work we have used the cosmological parameters $\rm H_o= 71 \, \rm km\,s^{-1}\,Mpc^{-1}$,\,$\Omega_m =0.27$,\, $\Omega_{\Lambda}=0.73$ to calculate distance.}

\section{OBSERVATION AND DATA REDUCTION}
We used archival data on Mrk~704 which was observed by
{\it{XMM-Newton}} on $2^{nd}$ November 2008 (obsID:0502091601) for a
total exposure of 98.2 ks. The EPIC-pn and MOS cameras were operated
in the small window mode using the thin filter. The data were processed
using SAS version 9 and the latest calibration database
as available on ${17}^{th}$ January 2010.

The data were first filtered using the standard filtering criterion. 
Examination of
the background rate above 10 keV showed that the observation was
partly affected by a flaring particle background after an elasped
time of $\sim87$ ks and this later period was therefore excluded to improve the signal-to-noise-ratio. We checked the photon pile-up using
the SAS task {\it epatplot} and found that there were no noticeable
pile-up in either EPIC-pn or MOS data. We quote results based on
EPIC-pn data due to its higher signal-to-noise compared to the MOS
data. We have used the good X-ray events (FLAG=0), pattern $\le$ 4. To
extract the source spectrum we used a circular region of
$45{\rm~arcsec}$, centred on the centroid of the source. We extracted the background spectrum from appropriate nearby circular
regions, free of sources. We created the ancillary response file
(ARF) and the redistribution matrix file (RMF) using the SAS tasks
{\it arfgen} and {\it rmfgen}.

We processed the RGS data using the SAS task {\it rgsproc}. We
assessed the background through examination of the light curve. We chose a region, CCD9, that is most susceptible to proton events
and generally records the least source events due to its location
close to the optical axis and extracted the background light
curve. We then generated a good time interval file  to filter the eventlist and extracted the first order source and background spectrum. The response matrices were generated using the task {\it
rgsrmfgen}.

\section{SPECTRAL ANALYSIS}

We used the Interactive Spectral Interpretation System (ISIS) version
1.5.0-20 for our spectral fitting. All uncertainties quoted are $90\%$
for one parameter of interest unless otherwise noted.

\subsection{The $0.3-10 \kev$ EPIC-pn spectrum}
We begin with the spectral analysis of the broadband ($0.3-10\kev$)
EPIC-pn spectral data. {We grouped the EPIC-pn data to a minimum of 300 counts per bin and used the $\chi^2$ statistics.} 
%The binning also complied with the optimal sampling theorem.} 
A single absorbed power-law model resulted in a poor fit to the pn
data, providing a minimum $\chi^2 = 11954$ for $559$ degrees of
freedom (dof). We therefore fitted the $2-10\kev$ data which was
devoid of any soft excess, with an absorbed power-law model.
 
The simple power-law model with absorption due to neutral hydrogen column in our Galaxy ({\tt wabs}) provided a
$\chi^2/{\rm dof}= 361/219$. The best fit neutral absorption column density $\nh \le 3.4 \times 10^{20} \rm cm^{-2}$ is consistent with Galactic column ($\nh^{\rm G} \,= \,2.97 \times \onlyten{20} \cmsqi$, \cite{2005A&A...440..775K}). So we fixed $\nh$ to this value. 
{ Previous studies on this source found strong intrinsic cold absorption (\cite{2008ApJ...673...96A}). We tested for the presence of intrinsic cold absorption by using the ({\tt zwabs}) component at the source redshift and found that the fit statistics did not improve, thus implying the absence of a fully covering intrinsic cold absorption.  We then tested for the presence of partial covering cold absorption by using the ({\tt zpcfabs}) model component. The fit improved by $\Delta {\chi}^2 = 64$ for two extra parameters. The best fit intrinsic neutral absorption column density is $\nh =  41^{+7}_{-8} \times 10^{22} \cmsqi$ and covering fraction is $ 0.18 ^{+0.04}_{-0.08}$.} Examination of the residuals showed a prominent
FeK$\alpha$ line at $\sim 6.4\kev$. Addition of a Gaussian line
profile improved the fit by $\Delta \chi^2 = 124$ for three additional
parameters, providing a statistically acceptable fit. The best-fit
parameters are: the photon index $\Gamma = 1.88 \pm 0.01$, iron line
centroid $E_{\rm FeK\alpha} = 6.39 \pm 0.03 \kev$, width $\sigma =
0.10^{+0.03}_{-0.03} \kev$
%line flux $f_{FeK\alpha} = 0.009 counts cm^{-2} s^{-1}$ 
and an equivalent width, $EW= 125 \pm 3 \ev$. The FWHM of the line corresponds to a velocity $v_{FWHM}\approx 11000 \kms$ which indicates that the line may originate from inner broadline region or the accretion disc. We also fit the line using {\it diskline} model which signifies line emission from a rotating disc. The best fit {\it diskline}  parameters are $E_{FeK\alpha} = 6.36 \pm 0.03 \kev$, norm=${2.58}_{-0.50}^{+0.62}\times 10^{-5}\punit $, the inner and outer radius $ R_{in}=10r_g$, $R_{out}=21000r_g$ ($r_g = GM_{BH}/c^2$ is the gravitational radius), emissivity index $\beta = -1.85$ and an inclination of $42.9 $ degrees. The observed $2-10\kev$ flux
is $ {1.10}_{-0.012}^{+0.011}\times 10^{-11} \funit$ corresponding to a
luminosity $L_{\rm X} = 2.18^{+0.02}_{-0.02} \times 10^{42}\lunit $. 

We extrapolated the best-fit $2-10\kev$ model ({\tt wabs $\times$ zpcfabs $\times$(powerlaw+gaussian)}) to the {soft X-ray} band (0.3 - 2$\kev$) and
found that there is a huge soft excess below $2\kev$. Figure
\ref{PN-excess} panel (f) shows the presence of soft excess over an absorbed
power-law.
The origin of soft excess in type 1 AGNs is still unclear. Several models such as single or multiple black-bodies, multicolor disk black-body, blurred reflection from partially ionised material, smeared absorption, and thermal Comptonization in an optically thick medium  can provide statistically good fit to the observed soft excess \citep{1998MNRAS.301..179M,2002MNRAS.331L..57F,2004MNRAS.349L...7G,2007ApJ...671.1284D}. Since we are mainly interested with the absorption phenomenon, any of the above models will serve our purpose to characterise the continuum in the soft X-ray band ($0.3-2 \kev$). We have  used the simple black-body ({\tt bbody}) to describe the soft excess emission. We added the {\tt bbody} component to the absorbed power-law plus Gaussian line model and performed the fitting in the $0.3-10\kev$ band. This fit resulted in an improvement,  $\Delta {\chi}^{2}= 2400$ for 2 extra parameters, compared to the absorbed power-law + Gaussian line model which was $\chi^2/{\rm dof}= 3659/554$. The black-body temperature is $\rm kT = 0.085 \pm 0.002 \kev$. The best-fit model is not statistically acceptable. So we further used one more {\tt bbody} and found an improvement in the fit by $\Delta {\chi}^{2}= 380$ for two extra parameters. The second black-body temperature is $\rm kT = 0.22 \pm 0.02 \kev$. 

Examination of the residuals showed broad absorption features in the $0.5-1 \kev $ region (see Fig.~\ref{PN-excess}, panel (d)). The absorption features likely include the \ion{O}{7} absorption edge and  Fe-M shell Unresolved Transition Arrays (UTA), which are the typical signatures of warm absorbers. Therefore, we created warm absorber table model {\it mtable} using CLOUDY to fit those absorption features.

\subsection{Cloudy modelling}

We created the model for warm absorption using the
photoionisation simulation code CLOUDY (version 08.00, last described by \cite{1998PASP..110..761F}). CLOUDY uses an extensive atomic database to predict the
absorption and emission spectrum from a cloud. The clouds are assumed
to have a uniform spherical distribution around the central source and
are photoionised by the source. CLOUDY performs the simulations by
dividing the surrounding medium into thin concentric shells referred
to as zones. The thickness of the zones are chosen small enough for
the physical conditions across them to be nearly constant. For each
zone the simulations are carried out by simultaneously solving the
equations that balance ionisation-neutralisation processes and
heating-cooling processes. The model predicts the absorptions and
emissions from such clouds in thermal and ionisation
equillibrium.  %The continuum used to generate the model using CLOUDY is described below.

We used an ionizing continuum of the central source that is typical of
AGNs and closely matches with the observed continuum in the X-ray
band.  This continuum is a power-law in the $1 \ev - 100 \kev$ band, and
another steeper power-law in the UV whose upper exponential cutoff is
parametrised with a temperature $T_{\rm BB}$ and the lower infrared cut
off by $T_{\rm IR}$. The continuum is expressed as

\begin{equation}
f_{\nu}={\nu}^{\alpha _{uv}} \exp(-h\nu /kT_{\rm BB}) \exp(-kT_{\rm IR}/h\nu)+\eta {\nu^{\alpha _{x}}},
\end{equation} 
where the coefficient $\eta$ is the relative normalisation between the
X-ray and the UV flux. It is calculated to produce the correct
$\alpha_{ox}$ for the case where the Big Bump does not contribute to
the emission at 2 keV and above.  Here $\alpha_{ox}$ is defined as the
slope of a nominal power-law connecting the continuum from 2500 $\AA$
and 2 keV, and is given by $\alpha_{ox}= 0.385 \log\left[{\frac{f_{\nu}(2500
\AA)}{f_{\nu}(2 keV)}}\right]$ \citep{1979ApJ...234L...9T}. The X-ray power-law
is only added for energies greater than 0.1 Rydbergs to prevent it from
extending into the infrared, where a power-law of this slope would
produce very strong free-free heating. The Big Bump component is
assumed to have an infrared exponential cutoff at k$T_{\rm IR}$= 0.01
Rydbergs. The free parameters in this model are $T_{\rm BB}, \alpha_{x},
\alpha_{ox}, \alpha_{uv}$ . 

{ The ionizing continuum to be used in CLOUDY should
represent the true continuum of Mrk~704. We used the best-fit
continuum model parameters derived from the broad-band EPIC-pn
data. However, the continuum parameters, derived without modeling the
warm absorption features, can be affected by the presence of warm
absorber components. Therefore we followed an iterative procedure. In
the first step, we obtained the X-ray continuum from the EPIC-pn
best-fit model without the warm absorbers. We used this X-ray
continuum and developed a CLOUDY warm absorber model. We fitted the
broad band EPIC-pn spectrum with this CLOUDY warm absorber model and
obtained new best-fit continuum parameters. In the second step we use
the new best-fit continuum parameters to create a more realistic warm
absorber model in CLOUDY. Finally we used this new warm absorber model
to fit the EPIC-pn spectrum.

We have obtained the best fit EPIC-pn continuum parameters without the
warm absorbers. In the first step we use them to model the warm
absorbers using CLOUDY.} The parameter values are $T_{\rm BB} =
9.8\times 10^5$ K ($\sim$0.085 keV), describing the soft-excess and
$\alpha_{x}= -0.88$, describing the X-ray powerlaw as derived from the
X-ray spectral fitting . The other parameter values were obtained from
radio-quiet AGN studies $\alpha_{ox}=-1.4$
\citep{1981ApJ...245..357Z,2009AAS...21348406Y} and $\alpha_{uv}=-0.5$
\citep{1994ApJS...95....1E}.

The CLOUDY table model was built using the methods described in
\cite{2006PASP..118..920P}.  We varied $\log {\xi}$ and $\log N_{\rm H}$ from
-3 to +3 and 18 to 24, respectively, and created a multiplicative
table model for the warm absorption. The electron density assumed was typical of BLR density $\sim 10^9 \rm \cmcubei$. The table model was subsequently imported to the ISIS
package.

\begin{figure}
  \centering
  \includegraphics[width=8cm,angle=-90]{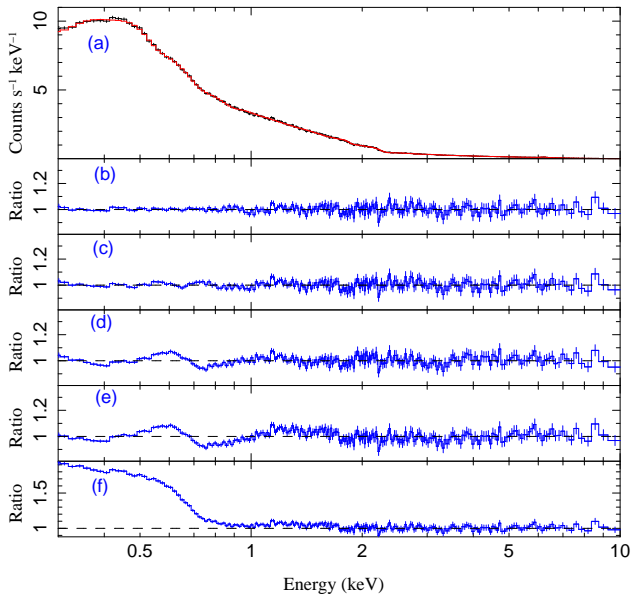}
  \caption{ The figure shows the improvement of the fit to EPIC-pn data on addition of different model components. The panel (a) shows the observed EPIC-pn data and the best fit model. The lower panels show the ratio of the data and the respective models. The panel (f) shows the presence of soft excess over a power-law. The panels (e) and (d) show the improvement in fit upon addition of a black-body component each of temperature 0.1 keV  and 0.25 keV respectively. The panels (c) and (b) show the improvement in fit on addition of a warm absorber component each of $\log \xi = 2.97$ and $\log \xi= 0.84$ respectively. Panel (b) also has a gaussian line fit in the soft X-ray. See Table \ref{PN-zxipcf} for details.}.
  \label{PN-excess}
\end{figure}

%\begin{figure}
  %\centering
  %\includegraphics[width=12cm,angle=-90]{fig7.ps}
  %\caption{The figure shows the best fit EPIC-pn spectrum in the 0.3-10 keV band. See Table \ref{PN-zxipcf} for details.}.
  %\label{PN-excess2}
%\end{figure}

First we used a single {\it mtable} component to model the absorption features.
% By a single component we mean a `cloud' along the line of sight having a certain physical state denoted by a specific set of parameters involving ${\xi}, N_H, n_e, r $. 
The model {\tt wabs $\times$ zpcfabs $\times$ mtable $\times$ (bbody(1)+ bbody(2) + powerlaw + gaussian)}
improved the fit ($\chi^2/{\rm dof} = 876/550$) by ${\Delta {\chi}^2} = 270$ for 3
extra parameters compared to the model without the warm absorber
component. The best fit parameters are $\log \xi$ = $0.99^{+0.13}_{-0.12}$ and $\nhwa$ =
${1.1}^{+0.12}_{-0.15}\times 10^{21} \cmsqi$. Further inspection of the residuals at
low energies showed additional absorption features in the $0.5-1 \kev$ range
that were not well described by the single warm absorber model. Therefore we used another {\it
mtable} component. The second component further improved the fit by
${\Delta {\chi}^2} = 40$ for 3 extra parameters resulting in $\chi^2/{\rm dof} = 566/544$. The best fit values
of the parameters are ${\log\xi}$ =$3.09 \pm 0.11$, $\nhwa$
=$1 \pm 0.13 $ $\times 10^{20} \cmsqi$ for the first component and
${\log\xi}$ =$0.88^{+0.08}_{-0.13}$, $\nhwa$ =$1.0 \pm 0.11$
$\times 10^{21} \cmsqi$ for the second component.  The power-law photon index
is $\Gamma = 1.865 \pm 0.055$. We also tested the presence of a third
warm absorber component but found no significant improvement 
and hence we excluded this third component from the analysis.

{Now we have the new best-fit continuum parameters
obtained from the continuum and warm absorber model fits to the
EPIC-pn data, we used the new continuum to create more realistic
warm absorption table model in CLOUDY.  We use this realistic CLOUDY {\it
mtable} model to fit the absorption features.}
% By a single component we mean a `cloud' along the line of sight having a certain physical state denoted by a specific set of parameters involving ${\xi}, N_H, n_e, r $. 
The model {\tt wabs $\times$ mtable $\times$ (bbody(1)+ bbody(2) +
powerlaw + gaussian)} improved the fit ($\chi^2/{\rm dof} = 876/550$)
by ${\Delta {\chi}^2} = 270$ for 3 extra parameters compared to the
model without the warm absorber component. The best fit parameters are
$\log \xi$ = $0.99^{+0.13}_{-0.12}$ and $\nhwa$ = ${1.1} \pm 0.15
\times 10^{20} \cmsqi$. Further inspection of the residuals at low
energies showed additional absorption features in the $0.5-1 \kev$
range that were not well described by the single warm absorber model
(see Fig.~\ref{PN-excess}, panel(c)). Therefore we used another {\it
mtable} component. The second component further improved the fit by
${\Delta {\chi}^2} = 36$ for 3 extra parameters resulting in
$\chi^2/{\rm dof} = 570/544$. The best fit values of the parameters
are ${\log\xi}$ =$2.97 \pm 0.11$, $\nhwa$ =$1^{+0.4}_{-0.11} $ $\times 10^{20}
\cmsqi$ for the first component and ${\log\xi}$
=$0.84^{+0.25}_{-0.12}$, $\nhwa$ =$9.7_{-1.7}^{+1.3}$ $\times 10^{20}
\cmsqi$ for the second component.

 In the soft X-ray range we found some positive residuals which could be the signature of an emission feature. We used a Gaussian line to the model the emission feature which improved the fit by ${\Delta {\chi}^2} = 11$ for two extra parameters. The Gaussian line $E= 0.565_{-0.009}^{+0.008} \kev$, flux = $1_{-0.11}^{+0.4} \times 10^{-4} \punit$, and the $\sigma = 0.01 \kev$. This unresolved line may be the OVII triplets. The observed
data, the best-fit two component warm absorbed model and residuals are
plotted in Fig.~\ref{PN-excess}, panels (a) and (b). { The total unabsorbed luminosity in 0.3-10$\kev$ $L_{\rm X} = 7^{+0.1}_{-0.06}\times 10^{42} \lunit $. The powerlaw luminosity and the soft-excess luminosities in the same band are $L_{\rm X} = 5.2^{+0.08}_{-0.05}\times 10^{42} \lunit $ and $L_{\rm X} = 1.78^{+0.02}_{-0.02} \times 10^{42}\lunit $, respectively. The blackbody with $\rm kT=0.085 \kev$ has a luminosity of $L_{\rm X}= 1.28_{-0.03}^{+0.03}\times 10^{42}\lunit $ and the blackbody with $\rm kT=0.22 \kev$ has a luminosity $L_{\rm X}= 4.8_{-0.25}^{+0.12}\times 10^{41} \lunit$. We list
the best fit parameters in Table \ref{PN-zxipcf}.} Due to the moderate spectral resolution,
individual narrow absorption and/or emission features are not
identified in the EPIC-pn data. However, the high resolution spectra
obtained are well suited for a detailed study of the warm absorption and emission.
Therefore, we have performed a detailed spectral analysis of the RGS data.

\subsection{RGS spectroscopy}

{\color {black}We grouped the RGS1 and RGS2 spectral data by a factor of two to ensure that each bin has non-zero counts.} We used {\it C Statistics} \citep{1979ApJ...228..939C} for analysing the RGS data as the $\chi^2$ statistics is not applicable in the case of low count data.

\begin{table*}

\begin{minipage}{180mm}
{\footnotesize

{%\scriptsize
\centering
  \caption{Details of the modelling of the emission lines in RGS spectra \label{lorentz}}
  \begin{center}
  \begin{tabular}{lllllllll} \hline\hline 

Gaussian & Rest & observed & outflow &Line flux & $\sigma$ & EW &($\Delta {\it C}$) \tablenotemark{a} & Ftest\\ 

 component & energy        &  Energy   & velocity    &        & & &&confidence  \\ 

           & ($\kev$) &  ($\kev$) & ($\kms$) & $\punit$ & $ (\ev)$ & $ (\ev)$ \\  \hline

{ O VII(r)} &  0.574 & ${0.5669}^{+0.0015}_{-0.0014}$ & ${5041}^{+640}_{-640} $&${3.86}^{+1.03}_{-2.28} \times {10^{-5}}$  &  ${1.3}^{+1.8}_{-0.09}$  & ${1.47}^{+1.22}_{-0.79}$ &  12 & 99 $\%$\\ \\
                              
{ O VII(i)} & 0.568 &${0.553}^{+0.004}_{-0.006}$& $777^{+2113}_{-3169}$ & ${2.83}^{+30.17}_{-1.36} \times{10^{-5}}$ & $0.9^{+1}_{-0.9}$ &$1.23^{+1.23}_{-0.59}$& 12 & $ 99 \%$\\ \\

{ O VII (f)} & 0.561 & ${0.5449}^{+0.0012}_{-0.0012}$& $ 90 ^{+500}_{-500}$ &${5.42}^{+4.5}_{-2.16}\times {10^{-5}}$  & $0.9^{+1}_{-0.9}$   & ${2.4}^{+1.8}_{-1.25}$ & 10 & $98 \%$ \\ \\ 

{ N VI(r)}$^{**}$ &  0.431 &${0.427}^{+0.0007}_{-0.0012}$& ${6300}^{+487}_{-835} $ & ${1.7}^{+1.1}_{-1.2}\times {10^{-5}}$  & ${1}_{-0.8}^{+0.9}$  & $ 0.51^{+0.65}_{-0.40}$& 6  & $ 89 \%$\\ \\

{ N VI(i)}$^{**}$   &  0.426 & ${0.4142}^{+0.0011}_{-0.0012}$& $249 \pm 704 $ &$ < 2.17 \times {10^{-5}}$  & ${0.25}^{+0.25}_{-0.25} $  &   --- & ---  \\ \\

 { N VI(f)}      & 0.420 & ${0.4085}^{+0.0005}_{-0.0005}$& $485^{+ 357}_{-357}$ &${2.30}^{+1.9}_{-2.1}\times {10^{-5}}$  & ${0.35}^{+1.8}_{-0.35} $  & ${0.65}^{+0.56}_{-0.44}$ & 6 &   $ 89 \%$ \\ \\

{ C VI}(Ly$\alpha$) &  0.367 &${0.3633}^{+0.0013}_{-0.0021}$& ${5670}^{+817}_{-1635}$ &${2.4}^{+3.5}_{-1.8}\times {10^{-5}}$ &  ${0.5}$  &  $ {0.49}^{+0.71}_{-0.36} $& 5 &  $ 83 \%$\\ \\ \hline \hline

\end{tabular} \\
\end{center}

\tablenotetext{a}{ The $\Delta {\it C}$ improvement quoted for each Gaussian fit is for an addition of 3 parameters}.
\tablenotetext{}{ All the parameters are obtained from simultaneous fit of RGS1 and RGS 2 data.}
 \tablenotetext{**}{ These lines are only found in RGS1 data. Therefore during simultaneous fit we could only give the upper limit of the flux of \ion {N}{6} (i) emission line and it is consistent within the error ranges for the same line detected in RGS1 alone. }

}
}
\end{minipage}

\end{table*}

 Both the RGS instruments operate in the
$0.35-2.5\kev$ band with one CCD inoperative in each case. We compared
the RGS1 and RGS2 data and found the two datasets agree well except
near $11.5{\rm~\AA}$. There was an emission feature at $11.5 \rm~\AA$
in the RGS2 data which was not present in the RGS1 data. This is likely an
instrumental feature and therefore we excluded the $11.3 - 11.7 \rm~\AA$
region from the RGS2 data and performed joint fitting of RGS1 and RGS2
data in the $7-38{\rm~\AA}$ band.

The presence of strong warm absorption features became more
evident in the high resolution RGS spectrum. We used a continuum model
similar to that inferred from the EPIC-pn data, i.e, the sum of a
power-law with $\Gamma = 1.865$ and two black-bodies with $\rm kT=0.085\kev $
and $0.22 \kev$ respectively. However we found that two black-bodies are
statistically not necessary in the RGS
spectrum. The higher temperature black-body has a low normalisation and hence does not improve the statistics.% starts from 0.35 keV and so unlike the EPIC-pn the lower energy soft excess does not affect the fitting. Hence

The continuum is best modelled by an absorbed power-law and a single
black-body. { The power-law photon index and the intrinsic neutral absorption component parameters were fixed to the best fit EPIC-pn values
since the RGS data alone cannot well constrain them.} However the power-law normalisation and the {\tt bbody}
parameters were left to vary. The equivalent neutral hydrogen column
density of the cold absorption was fixed to the Galactic value. On
fitting the spectrum with a power-law, a black-body, the Galactic absorption and a partially covered neutral
absorber we obtained a minimum ${\it C} = 3462$ with 3396 dof. The
best-fit {\tt bbody} temperature is $\rm  kT=0.0835^{+0.0020}_{-0.0012}\kev$.

With the continuum model defined we applied the warm absorption model
{\it mtable} and found an improvement in statistics by $\Delta {\it
C}=106 $ for 3 additional parameters. A close inspection of the
spectral residuals revealed further absorption features, seen as
negative residuals in the $16-20{\rm~\AA}$ band. Hence we applied
another component of the warm absorber. This second component further
reduced the value of C by $\Delta {\it C}=12$ for 3 additional
parameters ($C/{\rm dof} = 3344/3389$). The best-fit parameters are
${\log \xi}$ = $2.70_{- 0.15}^{+0.3}$ and ${\nhwa}$ =$2.69^{+1.11}_{-0.75}$
$\times 10^{20} \cmsqi $ for the first component and ${\log \xi}$
=$1.27^{+0.27}_{-0.52}$ and ${\nhwa}$ =$2 \pm 0.5 \times
10^{20} \cmsqi$ for the second component. 

After fitting the data using two components of warm absorbers, we
found emission lines as positive residuals in the spectra. There are
two triplets in the wavelength ranges $22-23{\rm \AA}$ and $28-29{\rm
\AA}$. We fitted the individual emission lines with Gaussian line
profiles. The best-fit line energies, widths and line fluxes are
listed in Table~\ref{lorentz}. We compared the observed line energies
and laboratory energies and identified them as the He-like triplet
emission lines of \ion{O}{7} in the $22-23{\rm \AA}$ band and
\ion{N}{6} triplets in the $28-29{\rm \AA}$ band. Such line triplets have also been detected in some of the Seyfert1 galaxies NGC~3783 \citep{2003ApJ...598..232B}, Mkn~335 \citep{2008A&A...484..311L}, NGC~4051
 \citep{2004MNRAS.350...10P, 2010A&A...515A..47N}.  These are the three
most intense lines of He-like ions: the {\it resonance} (w), {\it
intercombination} (x+y), {\it forbidden} (z). They correspond to
transitions between n=2 and n=1 states which are close in wavelength: 

\begin{enumerate}
\item  {\it resonance} line ({\it w}:  1$s^2 S^1_0$ -1s 2p $P^1_1$), 
 \item {\it intercombination} lines ( {\it x,y}:  1$s^2 S^1_0$- 1s2p $P^3_{2,1}$)
 \item {\it forbidden} line ({\it z}:  1$s^2 S^1_0$-1s2s$S^3_1$)
\end{enumerate}

The two intercombination lines ({\it x,y}) are not resolved and
detected as a single line in Mrk~704. The intercombination and the
resonant line of the \ion{N}{6} triplet were not prominent in the RGS2
data. However the $1\sigma$ upper confidence level of the line fluxes
derived from RGS2 lie well within the $1\sigma$ confidence interval
of the corresponding line fluxes derived from RGS1 data. The
$1\sigma$ upper limit on \ion{N}{6} ${(x+y)}$ line is
$1.2 \times 10^{-5} \punit $ and ${(\it w)}$ line
is $2 \times 10^{-5}\punit $ { obtained from RGS2,} while the
corresponding line flux obtained from RGS1 fit are
$1.9_{-1.9}^{+3.3} \times 10^{-5} \punit$ and
$4.8_{-4.2}^{+6} \times 10^{-5} \punit$,
respectively. Also we find that the lines are blue-shifted relative
to their rest wavelength. The resonant lines of both the triplets
arise from a cloud in outflow with a high velocity ($v\sim 5000
{\kms}$). While the outflow velocities of the other two
lines, (x+y) and z, of the triplets are very low. The velocities are
measured with respect to the systemic velocity. We checked
statistically whether we detect two distinct outflow velocities or whether one velocity can describe all the line shifts. We tied the
redshift of the \ion{O}{7} resonant line in the Gaussian model with the
other two lines of the triplet and found that the C statistic worsened
by $\Delta C = +8 $ from the fit where the resonant line was in a
faster outflow.

\begin{figure}[!h]
\begin{center}
\vbox{
   \includegraphics[width=6.5cm,angle=-90]{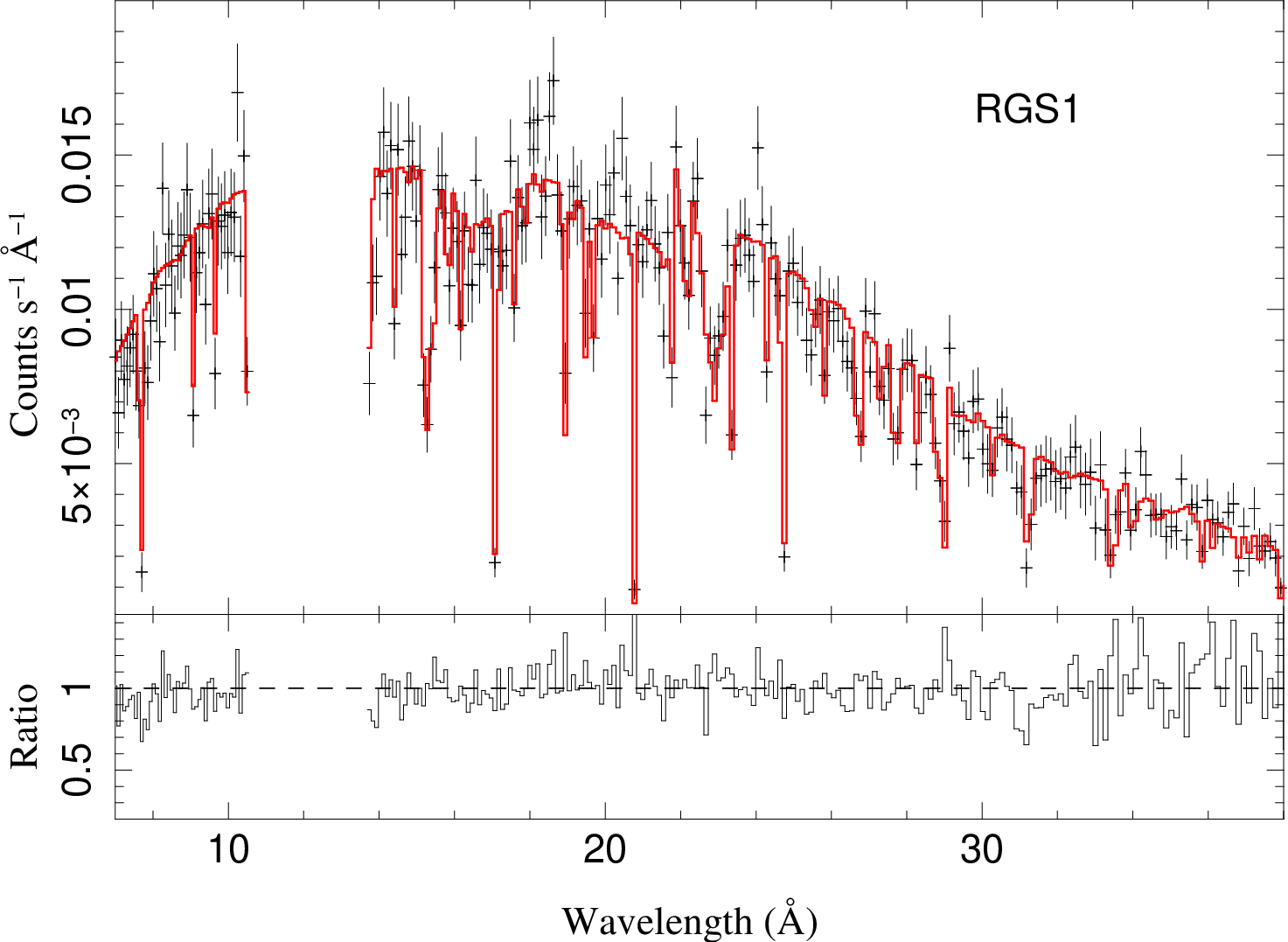} 
    
     \includegraphics[width=6.5cm,angle=-90]{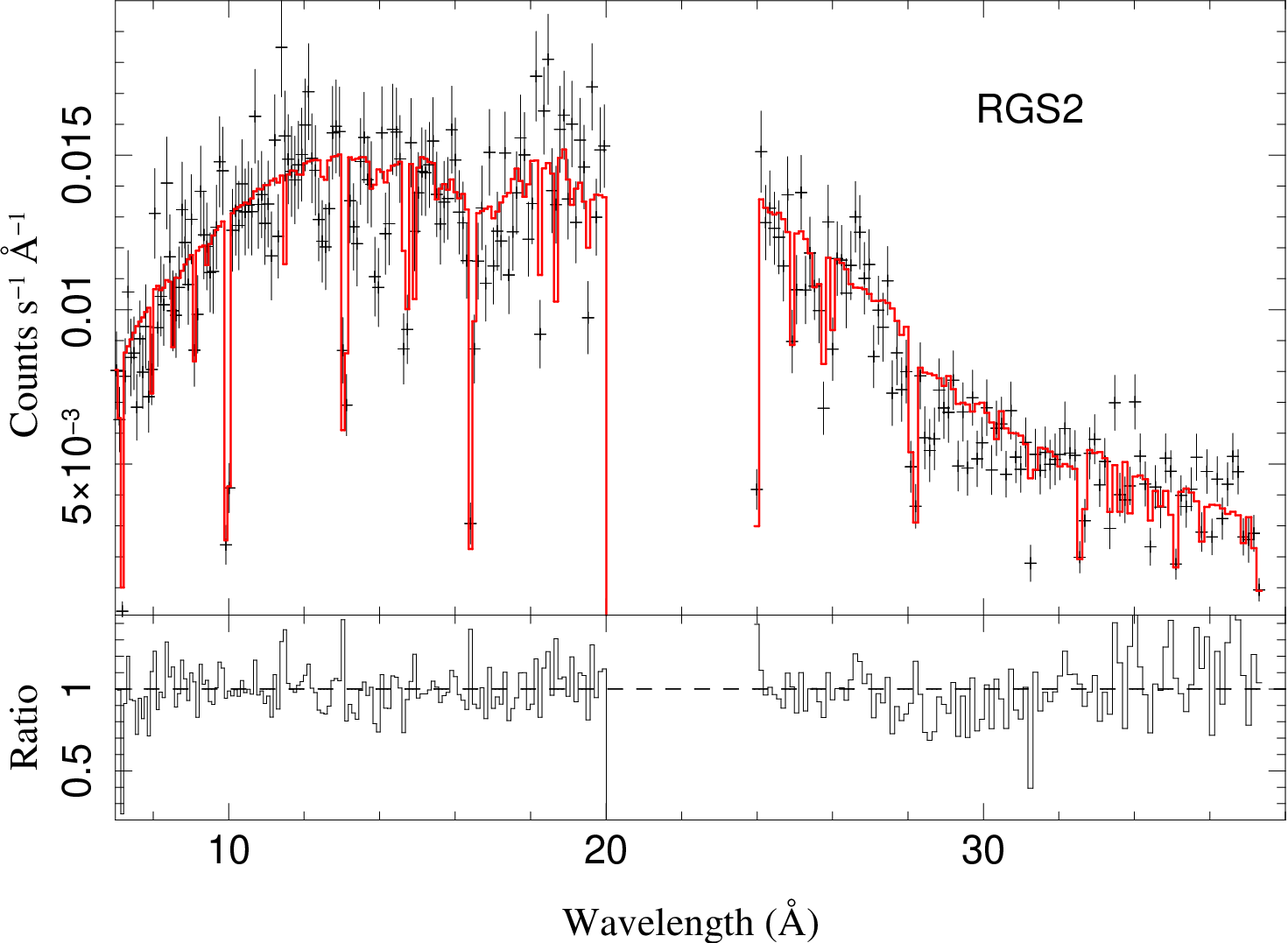}

       }
       \caption{ Shows the data with the best fit models for RGS1 and RGS2 along with the residuals.
       }
\label{RGS-full-fit}
\end{center}
\end{figure}

%\subsubsection{Monte Carlo simulation}

The detection of the \ion{O}{7} resonant line, which shows the signature of
a very high outflow velocity, is weak ($\Delta C \sim 12 $) as
inferred from the Gaussian fit. We therefore performed Monte Carlo
simulation for more rigorous test of the significance of the line. We
used the ISIS {\it fakeit} command to create 500 fake RGS1 and RGS2 data
sets using the best-fit model parameters of the actual dataset. The
faked data sets include source$+$background counts. We fitted the fake
data using the same model with background properly subtracted and
obtained the best fit Gaussian line parameters of \ion{O}{7}(r)
line. These parameter values (line energy and strength) were
distributed around their initial or the true values (the value used to
simulate the data). The true value and the error at the 90$\%$ confidence level on the line
energy distribution is 0.5669 keV and 0.001 keV, respectively, and for the line norm distribution is
$\ten{3.86}{-5} {\punit}$ and $\ten{2}{-5} {\punit}$ respectively. The confidence range of
detection is comparable to the confidence range obtained using the
C-statistic. Thus we detect the resonance line at a $2\sigma$ level. The detection of \ion{O}{7} intercombination and forbidden lines at similar significance at appropriate wavelengths suggests that these lines are highly unlikely to be spurious and we consider them as real features.

\begin{figure*}
  \centering
   \begin{center}
  \includegraphics[width=9cm,angle=-90]{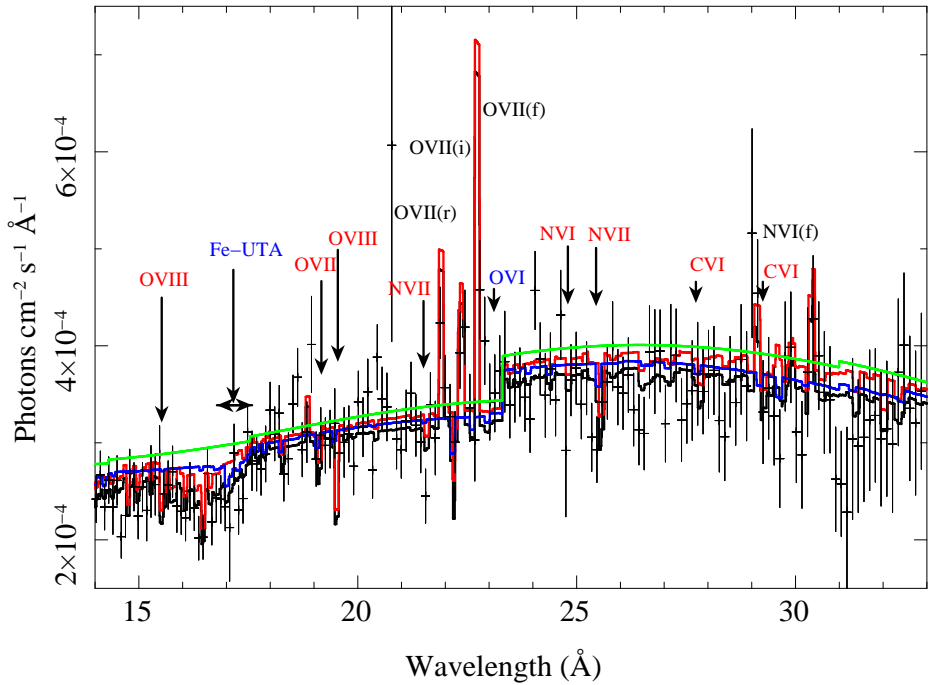}
   \end{center}
  \caption{The warm absorber model components obtained using Cloudy to jointly fit the RGS1 and RGS 2 data. The data had been grouped by a factor of 10 only for plotting purpose. For clarity we have shown only the RGS1 data. The green line indicates the continuum modified with Galactic absorption as well as the partially covering neutral absorber intrinsic to the source. The red line indicates the continuum modified by the higher ionisation phase of warm absorption $\log \xi =2.70$ apart from the Galactic absorption which models the \ion{O}{8}, \ion{O}{7}, \ion{N}{6}, \ion{N}{7}, \ion{C}{6} etc. The blue line indicates the continuum modified by the lower ionisation phase $\log \xi=1.27$ apart from Galactic absorption which fits the Fe M-shell UTA and OVI. The black line indicates the combined best fit. The emission lines are labelled in black letters.}.
  \label{diff-warm absorber-comp}
\end{figure*}

These He-like X-ray lines are of particular interest since their ratios are used as  plasma diagnostics. The Gaussian fit of these lines for both the triplets clearly indicates that the resonant line flux is comparable to the forbidden line flux within the errors (see Table \ref{lorentz}). Since the lines apparently arise from clouds having different outflow velocities, it is physically not meaningful to calculate the line ratios. The He-like emission line diagnostics (R and G ratios) enumerated in \cite{1973ApJ...186..327G} are based on the assumption that all the three lines of the triplet arise from the same cloud. The picture emerging in the case of Mrk 704 is that there are two distinct clouds where each cloud has its own predominating emission lines. We investigate the origin of these lines in more detail with CLOUDY modeling in the next section.
      
Apart from the two triplets we also detected \ion{C}{6} Lyman $\alpha$ but at a lower significance. The outflow velocity calculated from the line is same as that from resonant lines ($\sim$ 5000 km $s^{-1}$).

 As suggested by \cite{1993ApJ...411..594N} the warm absorbing clouds also emit. It is likely that there is a distribution of photoionised, warm clouds in all directions and the clouds along our line of sight give rise to the warm absorption while the clouds in all  directions contribute to the emission lines.  Thus, it is important to model both warm absorption and warm emission with a distribution of photoionised clouds. Hence we created a table model for warm emission using CLOUDY with the same input continuum used to model the warm absorbers. The electron density of the cloud assumed is $10^9 \cmcubei$.

Addition of a single warm emission component
improved the fit by $\Delta C= 26$ for 4 additional parameters
compared to the previous two component warm absorber model. We found
that only the intercombination and forbidden lines of \ion{O}{7}
triplet with similar outflow velocities (inferred from the Gaussian fit) are well fitted while the high
velocity resonant lines are not well described. From the study of line ratios of He like triplets, \cite{2000A&AS..143..495P} have shown that a weaker forbidden line is indicative of an emission from a  high temperature and high density plasma. So we created another warm emission model using CLOUDY with a higher density $\sim 10^{13} \cmcubei$ and used it to model the faster outflowing resonant lines of the triplets of Mrk~704 which have no forbidden line counterpart. The high density warm emission component fitted the resonant line of \ion{O}{7} and improved the fit by $\Delta C= 31$ for 4 extra parameters ($C/{\rm dof}= 3287/3382$). The best-fit parameters are: for the first component ${\log \xi}$ =$1.04_{-0.12}^{+0.4}$ and ${\nhwe}$ =$3.9\pm 0.1$ $\times 10^{20} \cmsqi$, and outflow velocity v= $ -30\pm 300\kms $; for the second component ${\log \xi}$=$1.14_{-0.4}^{+0.009}$ and ${\nhwe}$ =$ 1.0^{+0.007}_{-0.1}$ $\times 10^{18} \cmsqi $and outflow velocity v= $ 5190_{-300}^{+300}\kms $ {(The best fit parameters are listed as model 1 in Table ~\ref{RGS-cloudy}). The combined best fit-model as well as data  for individual RGS instruments are plotted in figure \ref{RGS-full-fit}.}  

Another possible scenario could be that we are not detecting the resonant line at all. It may be that a higher velocity outflowing cloud emits intercombination and forbidden lines which exactly coincides with the wavelength of the resonant and intercombination lines of the lower outflow velocity cloud. This possiblity is supported by the fact that the energy differences between (x+y), w and z lines are almost the same ($0.006 \kev$) within the errors ($\pm 0.001 \kev$). So keeping this in mind we tried to fit the line triplets using the same lower density ($\sim 10^{9} \cmcubei $) model, so that the forbidden lines are not suppressed. We used a second warm emission component of similar density ($\sim \onlyten{9} \cmcubei$) instead of the higher density model used in the previous fit. This improved the fit by $\Delta C= 29$ for 4 extra
parameters ($C/{\rm dof}= 3310/3380$) from the previous fit with a single warm emission component. The best-fit parameters of the second warm emitter component are :
${\log \xi}$ =$1.00^{+0.003}_{-0.05}$, ${\nhwe}$ =$ 2.8_{-0.5}^{+0.09}$
$\times 10^{18}\cmsqi$ and outflow velocity
$5000_{-300}^{+300}\kms $, where $\xi$ is expressed in units of
$\xiunit$ {(The best fit parameters are tabulated under model 2 in
Table~\ref{RGS-cloudy})}. The possible scenarios emerging from these analysis will be discussed in the next section.

\begin{table*}
\begin{minipage}{180mm}
{\footnotesize
\centering
 {%\scriptsize
  \caption{\scriptsize {Best-fit model parameters derived from the simultaneous fit of high resolution
    RGS 1 and RGS 2 observation of MRK-704(Using CLOUDY)} \label{RGS-cloudy}}
  \begin{tabular}{llllllll} \hline\hline

    Model  & parameters & Model 1\tablenotemark{a} & & & &Model 2\tablenotemark{b} \\ 
    components &            &                          &&&&                 \\ \hline

   Neutral absorption & $N_H$ ($10^{20}{\rm~cm^{-2}}$)   &
    $2.97$  &    &  &&$2.97$\\ \\

   Intrinsic neutral absorption & $N_H$ ($10^{22}{\rm~cm^{-2}}$)   &
    $50 $  &    &  &&$50$\\ 
    &Covering fraction & $0.22$ & & && $0.22$ \\ \\

    Power law  & $\Gamma$                     & $1.865$  &   & & &$1.865$\\
    & $norm_{\rm PL}$    & $(4.5\pm0.01)\times10^{-3}$&&&&$(4.5\pm0.01)\times10^{-3}$ \\ \\

    Blackbody  & $\rm kT_{\rm in}$ (eV)                & $83.5^{+2}_{-1}$ & & & &$83.5^{+2}_{-1}$  \\
    & $norm$     & $1.4_{-0.5}^{+0.7}\times 10^{-4}$  & & & &$1.4_{-0.5}^{+0.7}\times 10^{-4}$ \\ \\

    Warm absorber I &  $\nhwa$ ($10^{20}{\rm~cm^{-2}}$) & $2.69^{+1.11}_{-0.75}$ & & & & $2.69^{+1.11}_{-0.75}$\\
    & $\log \xi$   &  $2.7^{+0.3}_{-0.15}$ & $\Delta C=106 \tablenotemark{c}$ & & & $2.7^{+0.3}_{-0.15}$ &$\Delta C=106$\\
    & $(v/\kms)\, ^d$~  & $-540 \pm 100$ & & & & $-540 \pm 100$\\ & z &  $0.0274\pm {0.0004}$ & & & & $0.0274\pm{0.0004}$\\ \\

    Warm absorber II &  $\nhwa$ ($10^{20}{\rm~cm^{-2}}$) & $2_{-0.5}^{+0.5}$ &&& &$2_{-0.5}^{+0.5} $ \\
    & $\log \xi$   & $1.27_{-0.52}^{+0.27}$ & $\Delta C=12$ &&& $ 1.27_{-0.52}^{+0.27}$ & $\Delta C= 12$\\
    & ($v/\kms$)~  & $ -1350 \pm 300$ &&&&  $ -1350 \pm 300$ \\ & z & $0.0245\pm{0.001}$ &&&&  $0.0245\pm {0.001}$\\ \\

    Warm emitter I & $\nhwe$ ($10^{20}{\rm~cm^{-2}}$) & $3.9\pm{0.1} $ &&&& $3.9\pm{0.1} $ \\
    &  $\log \xi$   &  $1.04^{+0.4}_{-0.12}$ & $\Delta C=26$ &&&  $ 1.04^{+0.4}_{-0.12}$ & $\Delta C= 26$\\
    & ($v/\kms$)~  & $ -30\pm 300 $ &&&& $ -30\pm 300$ \\ & z &  $ 0.0289^{+0.0012}_{-0.002}$ &&&& $ 0.0289^{+0.0012}_{-0.002}$\\ 
    & norm                  &  $(5.3 \pm {2.5})\times 10^{-19}$ &&&&  $(5.3 \pm {2.5})\times 10^{-19}$\\ \\

    Warm emitter II & $\nhwe$ (${\rm~cm^{-2}}$) & $1.00^{+0.007}_{-0.11} \times 10^{18} $ &&&& $2.8^{+0.5}_{-0.09} \times 10^{18}$\\
    &  $\log \xi$   &  $1.14^{+0.009}_{-0.40}$ & $\Delta C^{*}= 31$&&&$1.00^{+0.003}_{-0.05}$  &$\Delta C= 29$ \\
    & ($v/\kms$)~  & $  -5190 \pm 300$ &&&&  $-5490 \pm 300$\\ & z &  $0.0119\pm{0.001} $ &&&& $0.0109\pm{0.001}$\\ 
    &norm             & $8.7^{+0.05}_{-0.4}\times 10^{-17} $ &&&& $ 1.57^{+0.05}_{-0.4}\times 10^{-19}$\\ \\

    $\rm C/dof$     &              & $  3287/3382$ &&&& $  3289/3382$ \\ \hline

\end{tabular}
\tablenotetext{a}{{\scriptsize This model is   {\tt wabs $\times$ cold absorber $\times$ warmabsorber(1)$\times$ warmabsorber(2)$\times$(powerlaw + bbody + warmemitter(1) + warmemitter(2))}, where the second warm emitter has a higher density($n_e=10^{13} \cmcubei$).}}

\tablenotetext{b}{{\scriptsize This model is same as above, except the second warm emitter has same density ($n_e=10^{9} \cmcubei$)}} 
\tablenotetext{c}{{\scriptsize The Reduction in C quoted as $\Delta C$ besides each model component indicates the improvement in statistics from its existing value on addition of that model.}} 
\tablenotetext{d}{{\scriptsize Outflow velocity.}} 
}
}
\end{minipage}
\end{table*}

\section {Discussion}

The  EPIC-pn spectrum above $2\kev$ 
is well described by an absorbed power-law with  $\Gamma\sim 1.86$, which is typical of Seyfert~1 galaxies. { Apart from Galactic absorption we detected an intrinsic cold absorber that covers the central X-ray source partially with a covering fraction of $\approx 0.22$. The low covering fraction is expected since the soft excess in the EPIC-pn spectrum is prominent and not altered significantly. The high column density ($N_H \sim 10^{23} \cmsqi$) of the partial covering absorber is suggestive of the fact that the cold absorption arises from the torus. This fits well with the picture that Mrk~704 is a polar scattered seyfert galaxy and hence the viewing angle is grazing the torus. 
%This partial obscuration may possibly be the reason for us to see emission lines from clouds that are farther out. 
The angular size of the central X-ray source as seen from the torus is likely $\sim 0.3\deg$ assuming a size of $\sim 50R_S$ for the central X-ray source and $\sim 0.1{\rm~parsec}$ for the torus. Thus the extended source can be covered partially by the torus.
%depending on the size of the X-ray emitting region. Hence partial obscuration is a possibility. The size of the X-ray emitting region of the accretion disc assumed is from $\sim 10 \, \rm to \, 100$ schwarzschild radius and the distance of the molecular torus $\sim 0.1 \, \rm pc$. 

There is a presence of a broad Fe-K$\alpha$ line at 6.4 keV with an
FWHM of $\sim 0.23 \kev$ corresponding to $v_{FWHM}\approx 11000 \kms$
. Hence the line can arise either from an accretion disk or inner
broad line region. The {\tt diskline} model fit also suggests that the
line arises from regions away from the inner accretion disk ($R =
10-20000r_g$). }
% 
%This model describes an emission from an accretion disc. The best fit outer radius of the accretion disk is $R_{out}=21000^{+10000}_{-16000}$  and inner radius $R_{in}= 10^{+100}_{-5}$ expressed in multiples of the schwarzchild radius and the emissivity index= -1.85 points to the fact that the Fe-K$\alpha$ line is more probably emitted from the inner broad line region.} 

The soft X-ray region 0.3-2 keV shows the presence of huge soft excess
over the power-law.The soft excess emission is described by two
black-body components with temperatures, $ \rm kT_{\rm BB} = 0.085$ and
$0.22 \kev$. The soft excess emission contributes $\sim 25 \%$ to the
$0.3-10\kev$ unabsorbed luminosity of $ 5.0^{+0.11}_{-0.06} \times 10^{42} \lunit$.  All
these features indicate that we are directly observing the emission
from the central AGN engine.

We estimated the bolometric luminosity $\lbol$ from the
monochromatic luminosity $L_\lambda(5100{\rm \AA}) = 4.9 \times
10^{39}{\lunita}$ (Peterson et al. 1998) and using
the relation in \cite{2000ApJ...533..631K},
\begin{equation}
 \lbol = 9\lambda L_{\lambda}(5000\AA),
\end{equation}
which gives $\lbol = \ten{2.2}{44} \lunit $. Thus $\sim 2.3\%$ of the bolometric luminosity is emitted in the X-ray band of $0.3-10\kev$. The
corresponding Eddington ratio can be obtained as $\lambdaedd =
\frac{\lbol}{\ledd} = 0.016 $, where $\ledd=1.38 \times 10^{38}
\frac{\mbh}{\msol} = \ten{1.13}{46} \lunit$, with blackhole mass,  $\mbh =\ten{8.2}{7} \msol$ estimated from the reverberation relation (see below).

\begin{table*}
\begin{minipage}{180mm}
{\footnotesize
\centering
  \caption{Details of the calculated warm absorber parameters of Mrk 704\label{warm absorberorber}}  \begin{center}
  \begin{tabular}{lllll} \hline\hline 

Calculated parameters & W.A component 1 & W.A component 2  & Units\\ 
                      & ($\log \xi$=2.7)   & ($\log \xi$=1.27)  \\ \hline \\

 outflow velocity & 540 $\pm$100 & 1350 $\pm$ 300  & km $s^{-1}$\\ \\

 $ r_{min}$           & $ R > 10^{-3}$ &  $ R > 2\times 10^{-3}$  & pc\\ \\

 $ r_{max}$           & $ R < 100 $  &  $ R < 10^3$  & pc\\ \\

 Volume filling factor($C_v$) & $ 3.81\times 10^{-3}$  & $ 2 \times 10^{-5}$  & --    \\ \\

Mass outflow rate    & $2.5 \times 10 ^{-5} $ & $5.0 \times 10 ^{-4}$  & $ \msolyi$  \\ \\

 $\frac{\rm L_{KE}}{\rm L_{ion}}$ & $10^{-6}$  & $10^{-5}$ & --\\ \\

 Width along LOS($\Delta R=\frac{\rm N_H}{\rm n_e}$) & $3 \times 10^{11}$ & $5 \times 10^{11}$ & cm \\ \hline \\ \\

\end{tabular} \\

\tablenotetext{}{ $ r_{min} $ and $ r_{max}$ are the minimum and maximum distance of the clouds from the central engine.}
\end{center}

}
\end{minipage} 
\end{table*}

\subsection {The warm absorbers }
We have found significant absorption of the soft X-ray emission by
partially ionised medium in Mrk~704. We have constrained the warm
absorption in Mrk~704 to consist of two strong absorber components as
is the case in other Seyfert 1 galaxies e.g., IRAS~13349+2438 \citep{2001A&A...365L.168S}, NGC~3783 \citep{2003ApJ...597..832K}, Mrk~841 \citep{2010A&A...510A..92L}. The lower ionisation phase
($ \log \xi$=1.27) contributes to the Fe M shell UTA and \ion{O}{6}, while the
higher ionisation phase ($\log \xi$=2.7) gives rise to the absorption
from \ion{O}{7}, \ion{O}{8}, \ion{N}{6}, \ion{N}{7}, and \ion{C}{6} {(see Fig.~\ref{diff-warm absorber-comp})}. The broad range of charge states observed
along the line of sight is typical of Seyfert outflows
\citep{2003ApJ...598..232B}.

The lower ionisation phase is found in outflow with a velocity of
$\sim 1350 \kms$ , while the higher ionisation phase has a
lower outflow velocity of $\sim 540 \kms$. The picture is
consistent with a radiatively driven outflow, where the lower
ionisation phase has a greater optical depth for the ionising
radiation and hence the cloud acquires more momentum from the ionising
flux than the higher ionisation phase. Another possible
explanation is that the high velocity, low ionisation component
originates closer to the nucleus where the escape
velocity is very high, and the high ionisation component could be
launched much further away. In keeping with the Polar scattered
Seyfert scenario it is tempting to think of the higher ionisation (lower velocity)
phase to emanate from the torus (torus wind). However this is unlikely
since for such a highly photoionised plasma to exist at such
a large distance from the nucleus, the corresponding density of the cloud has to be very
low.

%high ($\sim 10^{7}\rm cm^{-3}$) which is inconsistent with the density of the torus.\citep{2004ApJ...600L..35M,2001ApJ...557....2C }.

\subsubsection{Energetics of Warm absorbers}
As in many Seyfert 1 galaxies, the warm absorbers in Mrk~704 are in
outflow. The associated mass outflow rate can be estimated by assuming
that the warm absorber is in the form of spherical outflow with the
gas density smoothly decreasing as $1/r^2$ and with a uniform volume
filling factor $C_v$. Following \cite{2005A&A...431..111B}, the mass
outflow rate can be written as
  
\begin{equation}
\dot{M}_{\rm out}  \sim \frac{1.23m_{p} L_{\rm ion} C_{v} v \Omega}{\xi},
\label{mdot}
\end{equation}
where $L_{\rm ion}$ is the ionizing luminosity, $v$ is outflow velocity and $\Omega$ is the solid angle subtended by the outflow.
The covering fraction is given by
\begin{equation}
C_v \sim \frac{({L_{\rm abs}} + {L_{\rm ion}(1-e^{-\tau})}) \xi}{1.23 m_p c L_{\rm ion} \Omega v^2},
\label{cv}
\end{equation}
where $\tau$ is the Thomson scattering optical depth and ${L_{\rm abs}}$ is the luminosity absorbed by the cloud in the whole energy band of $0.3-10 \kev$.

We have calculated ${L_{\rm abs}}$ for the two warm absorber phases
from the best-fitting models to the RGS data. The luminosity absorbed
by the higher and lower ionisation phases is ${L_{\rm abs}}=$ $1 \times
10^{41}{\lunit}$ and $\ten{4.0}{41}{\lunit}$
respectively, in the $0.3-10\kev$ band. We calculated the ionising
luminosity as the unabsorbed luminosity in $0.3-10 \rm keV$ band, and the Thomson scattering optical depth using the best-fit
absorption columns to be $\tau \sim \ten{1.7}{-4}$ and $\sim \ten{1.3}{-4}$, respectively, for the higher and lower ionisation
components.  Using the above values and Eqn.~\ref{cv}, we obtained  volume filling factors $C_v$ of $\sim \ten{4.8}{-3}$ and
$\sim 4.8\times 10^{-4} $ for the higher ionisation and lower
ionisation phases, respectively. We have used the outflow velocities
derived from the best-fitting warm absorber model to RGS data.  Using
Eqn.~\ref{mdot}, we estimate the mass outflow rate, $\dot{M}_{\rm out}
\sim \ten{2.5}{-5} \msolyi$ for the high ionisation
phase and $\dot{M}_{\rm out} \sim \ten{5.0}{-4} \msolyi$
for the low ionisation phase.

 We used $\Omega \sim 1.5$, typical for Seyfert 1s \cite{2005A&A...431..111B}. The results are listed in Table \ref{warm absorberorber}.  For comparison, we have also estimated the mass accretion rate for Mrk~704 as
\begin{equation}
\dot{M}_{\rm acc} = \frac{\lbol}{\eta c^2},
\end{equation}
where we assumed the accretion efficiency $\eta = 0.1$. The mass accretion rate thus estimated is $\dot{M}_{\rm acc} = 0.038 \msolyi$. Thus the total  mass outflow rate in the warm absorbers is $13\%$ of the mass accretion rate.
 
\subsubsection{Location of the warm absorbers}
The maximum distance of the warm absorbing clouds can be estimated using the argument that the size of the cloud $\Delta R$ is very less than its distance from the source R. In that case using the expression $\xi=\rm L/nR^{2}$ we have\\

\begin{equation}
{\frac{\xi R N_H}{L} < 1},
\end{equation}
which gives $R < 100$ pc and $R < 10^{4}$ pc for the high ionisation and low ionisation components respectively. The minimum distance of the cloud can be estimated by noting that its outflow velocity must be greater than or equal to the escape velocity at that radius.

\begin{equation}
\frac{GM}{R} < \frac{v^{2}}{2},
\end{equation}
From this we get $R > 10^{-1}$ pc and $R > 10^{-3}$ pc for the high ionisation and low ionisation components respectively. Recently \cite{2009arXiv0903.5310S} and \cite{2007ApJ...659.1022K} have considered more realistic approaches to constrain the warm absorber distances.

 The mass of the central blackhole was calculated using an emperical relation obtained by \cite{2000ApJ...533..631K} using the reverberation techniques,

\begin{equation}
M = 1.464 \times 10^5 \left[\frac{R_{\rm BLR}}{\rm lt-days}\right] \left[ \frac{v_{\rm FHWM}}{\onlyten{3}\kms} \right]^2 M_{\odot},
\label{mass-lum}
\end{equation}

\begin{equation}
R_{BLR}=(32.9^{+2.0}_{-1.9}) \left[ \frac{\lambda L_{\lambda}(5100 \AA)}{\onlyten{44} \lunit} \right]^{0.700\pm0.03} \rm lt-days,
\label{R-lum}
\end{equation}
where $R_{BLR}$ is the distance of the broad line region from the
central engine which is estimated in terms of time lags in Balmer
emission lines. $ v_{\rm FHWM}$ is the velocity broadening of the
Balmer lines assuming that the line widths represent a Keplerian
velocity. The value of $\lambda L_{\lambda}(5100 \AA) $ in case of
Mrk 704 was obtained from previous optical studies by
\cite{1998ApJ...501...82P}. The ${v_{\rm FHWM}}$ of the $H_{\alpha}$ line
is $\sim 6750 \kms$ and was obtained from the paper
\cite{1996ApJ...470..280M}. The estimated mass turns out to be $\sim
\ten{8.2}{7} \msol$.

\subsection{ The warm emitters}

Apart from warm absorption, we detected emission features arising from
He-like ions of O and N and H-like ion of C. 
Simple Gaussian profile
fits suggested that the resonance lines are significantly blueshifted
by $v \sim 5000\kms$ relative to the forbidden and
intercombination lines. This is possible if two sets of kinematically
distinct clouds contribute to the emission line triplets such that
resonance lines predominantly arise from high velocity clouds while
the forbidden and intercombination lines arise from the low velocity
clouds. The forbidden lines from the high velocity clouds can be
suppressed if the density is high or the UV radiation field is
intense.  The density dependence arises as the $^3S_1$ level electrons can be
collsionally excited to the $^3P$ levels. At sufficiently high density, more than the critical
density for the forbidden line ($\approx \ten{3}{10} \cmcubei$ for \ion{O}{7},\, \citep{2001A&A...367..282N}),
the collisional excitation from $^3S_1 \rightarrow {^3P}$ weakens the forbidden line. 

The He-like line ratios can also be
altered by the presence of UV radiation \citep{1969Natur.221..947G}. If
the photon flux is high at the wavelength corresponding to the energy
difference between forbidden and intercombination line levels, i.e
$1900 \rm \AA$ for \ion{N}{6} and $1630 \rm \AA$ for \ion{O}{7} ($6.5 \ev$ and $7.6 \ev$
respectively), the $^3S_1$ level can be photoexcited to the $^3P$
levels. The more the photon flux at these energies, the more is the
forbidden level depleted to intercombination level (2 $S^3 \rightarrow
2 P^3$). Thus the forbidden line becomes weaker and the
intercombination lines stronger.  

 In keeping with the polar scattered Seyfert scenario we can say that the clouds that are in the shadow of the torus face a weaker UV flux and hence emit predominantly forbidden and intercombination lines. They are also of lower densities. But the higher density clouds facing the central engine have their forbidden lines supressed and have prominent resonant lines.

 The widths of the emission lines are $\leq 700 \kms$,
similar to the emission lines observed from optical narrow-line region
(NLR). The density in the NLR is typically $n_e \sim 10^3-10^4 \cmcubei$. At these densities (i.e below the critical densities) the forbidden lines of \ion{O}{7} are expected to be strong \citep{2002astro.ph..3021K}. Thus, it is possible that the forbidden and intercombination lines arise from the optical NLR. The high-velocity resonance lines arise from a more dense region nearer the central engine.

\begin{figure}[!h]
\vbox{
   \includegraphics[width=5cm,angle=-90]{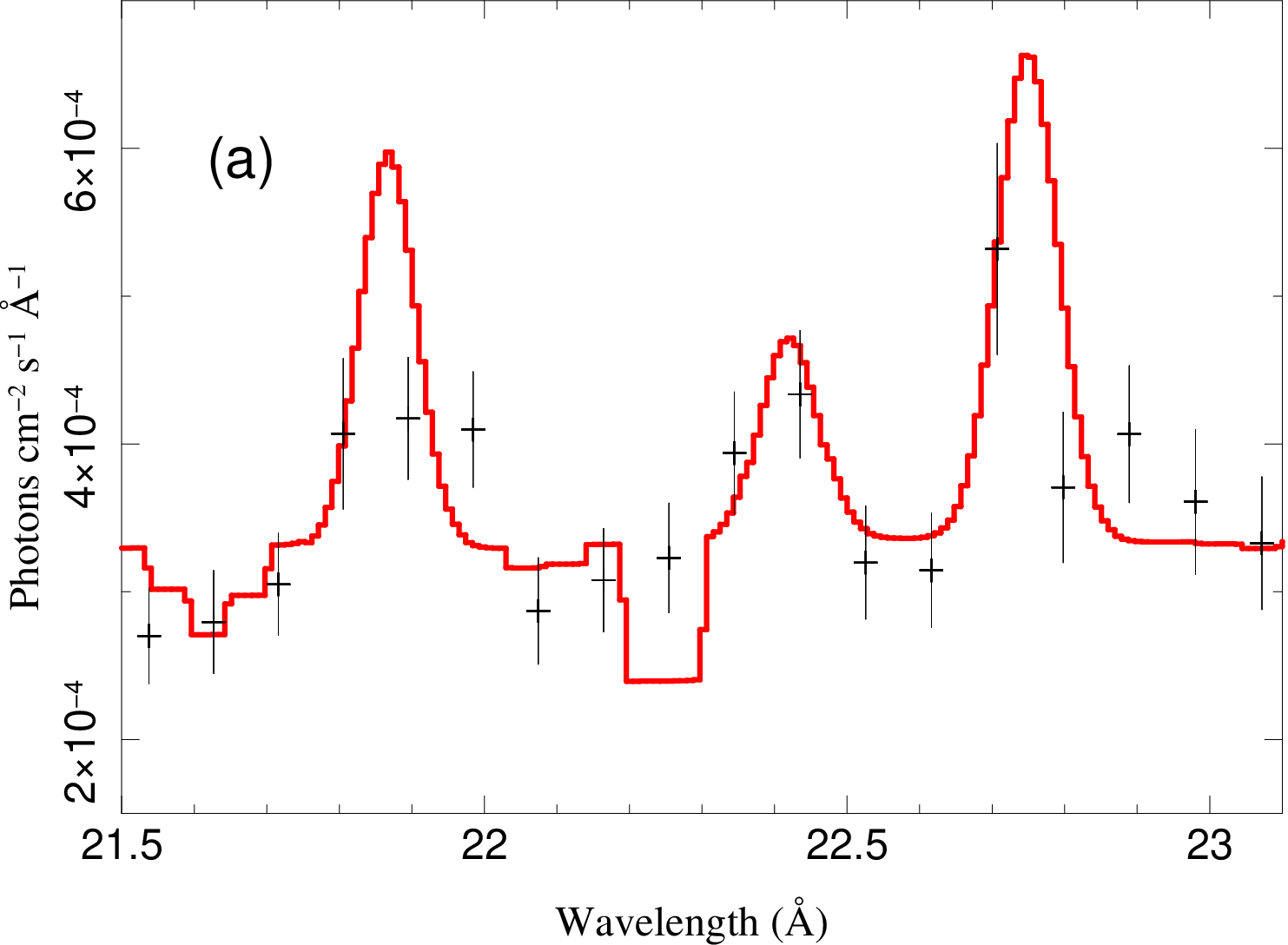} 
    
     \includegraphics[width=5cm,angle=-90]{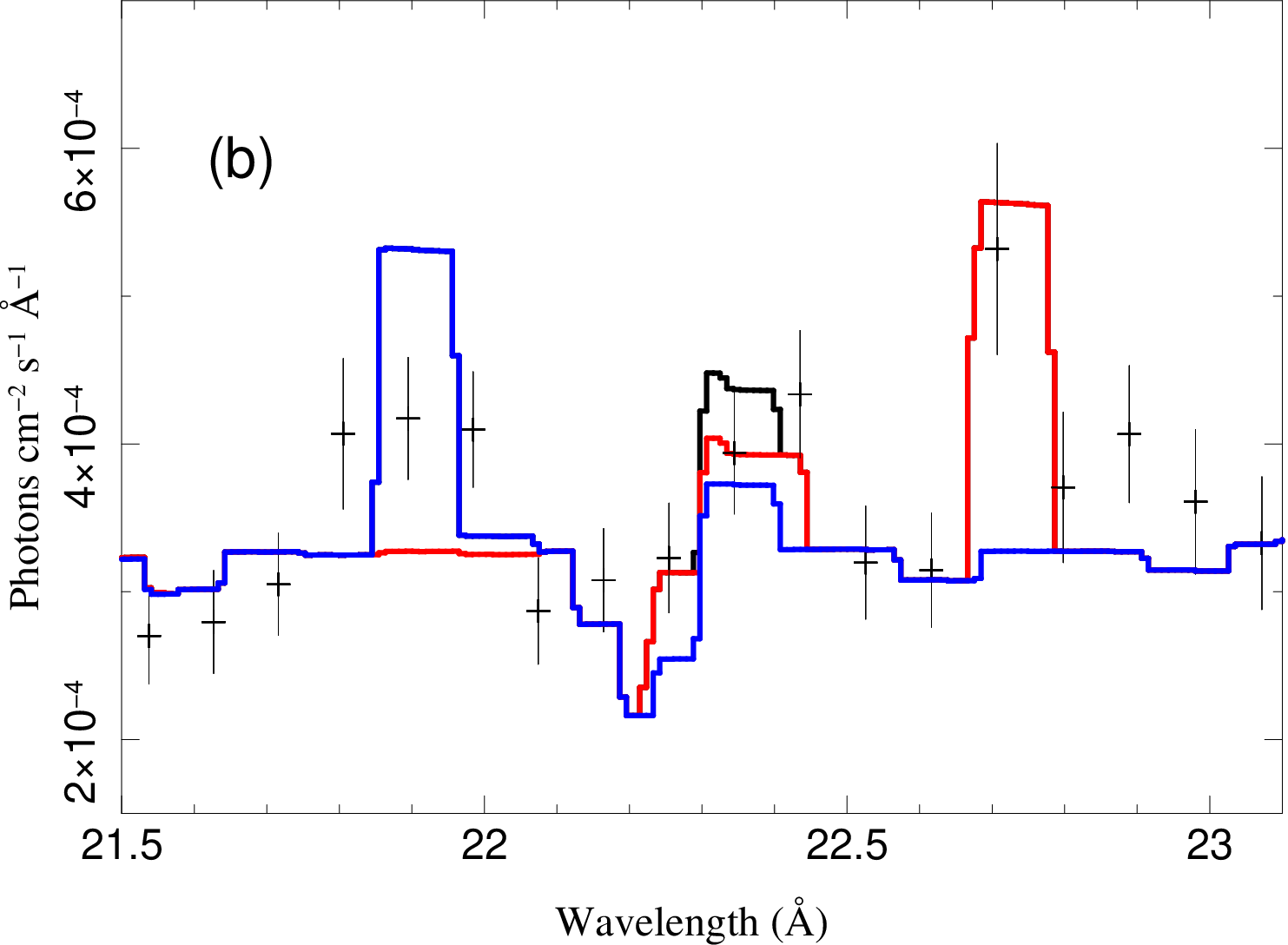}
     \includegraphics[width=5cm,angle=-90]{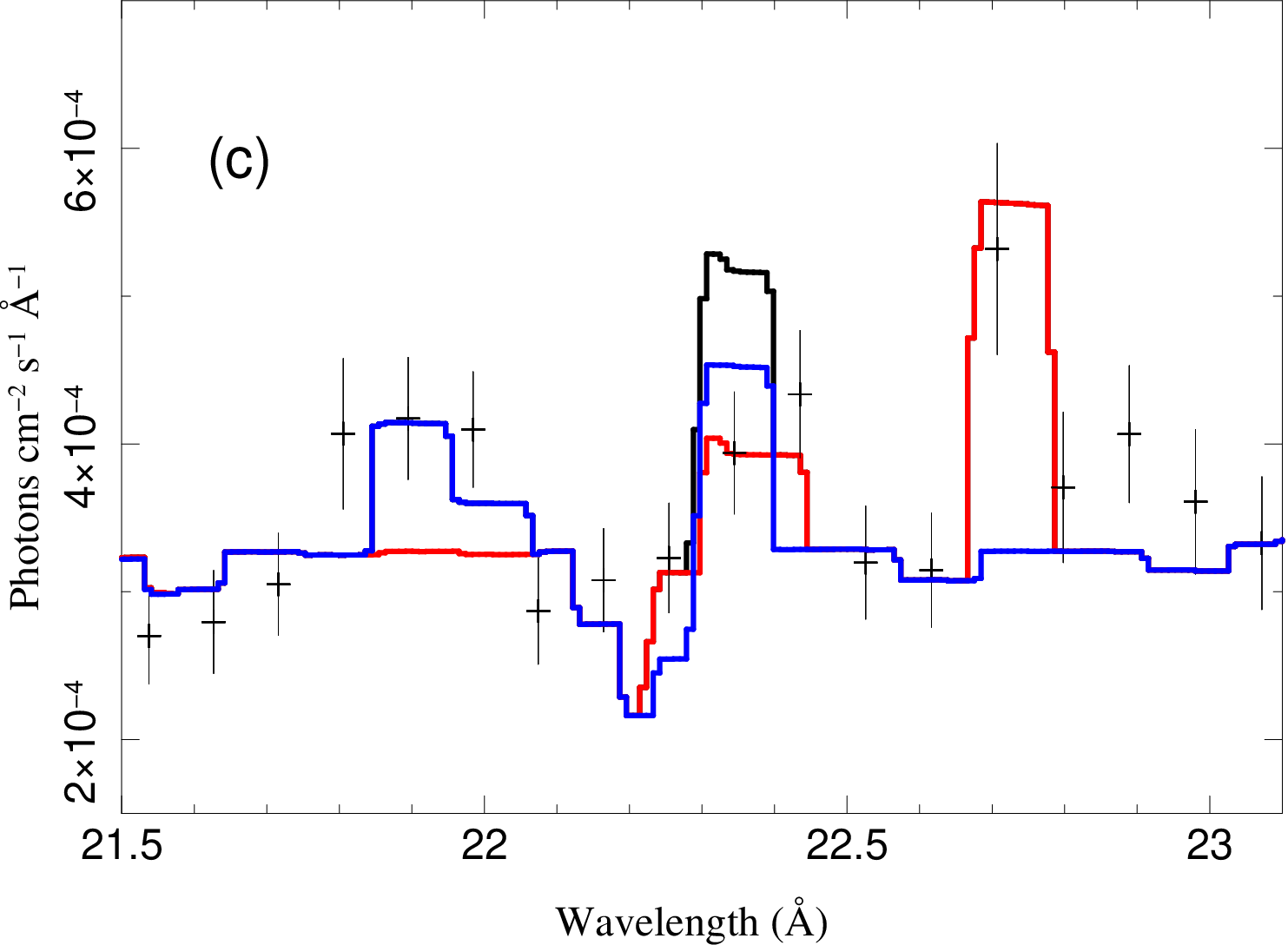}

       }
       \caption{ Shows the modelled \ion {O}{7} emission lines. {\it The panel (a)} shows Gaussian fit to the three lines. The {\it panel (b)} shows the fit using two warm emission models created using CLOUDY. One is created using a cloud density of $10^{13} \cmcubei$ and the other using a density $\onlyten{9} \cmcubei$. The blue line shows the higher density model for the faster outflowing resonant line while the red line shows the lower density model for the forbidden and intercombination lines. The {\it panel (c)} shows the fit using the emission line models created
       using a cloud density of $\onlyten{9} \cmcubei$. The blue line models
       the faster outflowing emission lines while the red line models
       the lower velocity cloud emission lines. In the {\it panels (b) and (c)} the black line shows the combined 
       best fit.
       }
\label{OVII}
\end{figure}

Another possibility is that we do not really detect the resonance lines but detect the intercombination and the forbidden lines. If one set of emission lines is blueshifted such that the intercombination line of the low velocity cloud concides with the forbidden line of the high veolcity cloud, then the overall line structure will be triplet - the forbidden line from the low velocity cloud ($0.545 \kev$), the blended intercombination line from the low velocity cloud and the forbidden line from the high velocity cloud ($0.553 \kev$), and the intercombination line from the high velocity cloud ($0.5669 \kev$). The energy difference between those lines (x+y), w and z are similar within the errors and hence this overlapping is a possibility.  In this case, emission lines from two low density clouds with different outflow velocities describe the observed triplet lines. However, this model over predicts the central intercombination line (see Fig \ref{OVII}, panel c) and is marginally statistically poorer than the previous high density fit.

However, there are a few caveats in our analysis. Emission from any cloud
is almost an isotropic process and the viewing solid angle subtended
at our eye by these clouds being very less, the emission features are
mostly very weak.  Also the emission features are contributed from all
clouds distributed around the source unlike the absorbers which lie
only along our line of sight to the source. So the physical conditions
in such emitting clouds are varied and complex. Hence fitting the
emission lines by discrete parameter values may not give us a clear
picture. As pointed out by \cite{2002astro.ph..3021K} , the emission
features could be modelled by a continuos distribution of $\xi$ and
$N_H$ as against our discrete cloud assumption. Each cloud with
different physical conditions might be contributing little to the
spectrum and what we see is the sum of the effect of all such
clouds. The electron densities $n_e$ assumed while modelling with
Cloudy were $10^9 \cmcubei$ and $10^{13} \cmcubei$ which might not be appropriate for clouds with widely varying conditions.

{The observations of the emission lines from this type 1 AGN is likely due to the reduced X-ray continuum by the partial covering neutral absorber as revealed in the EPIC-pn data and earlier by \asca{} and \swift{} observations \citep[see][]{2008ApJ...673...96A}. This implies that the cold partial covering absorber is closer to the nuclear X-ray source than the X-ray line emitting region. If the cold absorption is provided by the torus, then the X-ray line emitting region must be outside this torus. This suggests that the low velocity X-ray  line emitting region is similar to the X-ray narrow line region observed from Seyfert 2 galaxies such as NGC~1068 \citep{2002astro.ph..3021K} }

\section{Conclusion}
We have performed a detailed analysis of a long \xmm{} observation of
Mrk~704. The main results are as follows:

\begin{enumerate}
\item The 0.3-10 keV continuum is well described by a power-law with
  slope $\Gamma \sim 1.865$ and a soft X-ray excess component
  consisting of two black-bodies with temperatures $\rm kT\approx 0.085$ and
  $0.22\kev$. The soft excess contributes $\sim 25\%$ to the
  $0.3-10\kev$ intrinsic flux.

\item { We detected the presence of a partially covering cold absorber intrinsic to the source. The high column density {($N_H \sim 10^{23} \cmsqi) $} of the cold absorber supports the fact that Mrk 704 is a polar scattered Seyfert galaxy and we are looking at the source along a line of sight that grazes the torus.}

\item A broad FeK$\alpha$ line is detected at $6.4 \kev$ with $v_{\rm FWHM}\approx 11000 \kms$. The line can arise either from the inner broad line region or the outer regions of the accretion disk.

\item A two phase warm absorber with $  \log \xi\approx 1.27 $ and
  $  \log \xi\approx 2.7$ is detected in the spectra. The lower
  ionisation phase has a faster outflow velocity $\sim 1350 \kms $ while the higher ionisation phase has a lower velocity
  $\sim 450 \kms$.

\item Weak emission features of He-like triplets of \ion{O}{7} and
  \ion{N}{6} are identified. One of the lines in the triplets is in
  much faster outflow ($\sim 5000 \kms$) than the other two lines
  which are from a cloud with outflow velocity consistent with zero.
  The CLOUDY modelling points out to two possible situations. The
  first is that there are two sets of clouds with two different
  velocities and two different densities ($n=10^9$ and $10^{13}
  \cmcubei$). The higher density, higher outflow velocity cloud
  produces the resonant line while the lower density lower velocity
  cloud gives rise to the intercombination and forbidden lines. The
  second situation is that there are two clouds of low density ($n=
  10^9 \cmcubei$) but with different outflow velocities giving only
  intercombination and forbidden lines.

\item The low density emission phase is likely similar to the X-ray
  narrow-line region observed from many Seyfert~2 galaxies. The
  high velocity, outflowing emission phase is unique to Mrk~704.
\end{enumerate}

$Acknowledgements:$ We thank an anonymous referee for his/her
insightful comments and suggestions. We also thank Smita Mathur,
Matteo Guainazzi for useful discussions and comments on the paper. We are also thankful to Gary Ferland and Ryan Porter for discussions on the CLOUDY modeling. One of the authors SL is grateful to CSIR, Government of
India for supporting this work.

\bibliographystyle{apj}
\bibliography{mybib}

%***************************************************************************************************
%**************************************************************************************************

%\begin{figure}
 % \centering
  %\includegraphics[width=18cm,height=18cm,angle=-90]{fig6.ps}
  %\caption{ The combined best fit RGS1 and RGS2 spectrum. The data from 11.3 $\rm \AA$- 11.7 $\rm \AA$ in RGS2 were excluded while fitting.} %For identifying the emission and absorption features please see figure \ref{diff-warm absorber-comp} .}.
  %\label{RGS-full-fit}
%\end{figure}

%\newpage
%\begin{landscape}
\begin{sidewaystable*}
%\begin{table*}
{
\footnotesize
\centering
  \caption{Improvement in statistics on addition of a model as derived from the medium resolution
    EPIC-PN data analysis of MRK-704 \label{PN-zxipcf}}
\begin{center}
  \begin{tabular}{lllllllllll} \hline\hline 
    
 Models\tablenotemark{d}   &  PL \tablenotemark{a}       & wabs\tablenotemark{a}  & PCA\tablenotemark{a} & $\rm GL_{_{FeK\alpha}}\tablenotemark{a} $  & BB(1)\tablenotemark{a} & BB(2)\tablenotemark{a}      & WA(1)\tablenotemark{a} & WA(2)\tablenotemark{a} &  $\rm GL$ &  $ \rm \chi^2/dof$ \\ \hline

 & ($\Gamma$)   & $\rm N_{H}^{Gal}$  &  $\rm N_{H}^{PCA}$ & $\rm E \, ,\,\, \sigma $           & $\rm kT$ & $\rm kT $&  $\rm \nhwa$ & $\rm \nhwa $ &    $\rm E \,, \, \, \sigma$ &   \\

 &              &     $ (10^{20} \cmsqi)$        &      $ (10^{23} \cmsqi)$           & $(\rm keV)$        & $(\rm keV)$         &   $(\rm keV)$      & $(10^{21} \cmsqi)$ &  $(10^{20} \cmsqi)$  &  $(\rm keV)$ &  \\

 &              &              & $\rm C_{v}$ & $\rm Flux\tablenotemark{e} $        &          &         &  $\log {\frac{\xi}{\xiunit}}$  &  $\log {\frac{\xi}{\xiunit}}$  & $\rm Flux  \tablenotemark{e}$  &  \\ \hline

 %&  & $(10^{20} cm^{-2})$  &  & $\rm Gauss_{FeK\alpha}$ & BB(1) & BB(2) & WA(1) & WA(2) &  $\rm Gauss$ &  $ \rm \chi^2/dof$\\ \hline \hline \\ 

%------------------------------------------------------------------------------------------------

 Model 1 $^{b}$   & $1.76$ & $ 2.97$ & -- & -- & -- & -- & -- & -- & -- & 361/219 \\ 

  Model 2 $^{b}$ & $1.88$ & $2.97$     & $ 4.4$&--&--&--&--&--&--& 297/217  \\

                       &                 &                                    &  $0.31 $ &            \\

 Model 3 $^{b}$ & $1.88$ & $ 2.97$ & $ 4.4$ & $ 6.39 \, ,\,\,0.1 $ &--&--&--&--&--& 173/214                   \\

                       &                 &                                    &  $ 0.31 $ & $ 1.7 \times 10^{-5}$           \\ \hline

%------------------------------------------------------------------------------------------------------

Model 4  $^{c}$ & $1.88$ & $ 2.97$ & $ 7.0$ & $ 6.39 \, ,\,\,0.1 $ & $  0.089$&--&--&--&--& 1259/552                   \\

                       &                 &                                    &  $ 0.21 $ & $ 1.7 \times 10^{-5} $           \\

Model 5  $^{c}$ & $1.88$ & $2.97$ & $ 6.1$ & $ 6.39 \, ,\,\,0.1$ & $ 0.089 $& $0.3 $ &--&--&--& 876/550                   \\

                       &                 &                                    &  $ 0.21 $ & $  1.7 \times 10^{-5}$           \\

Model 6  $^{c}$ & $1.85$ & $ 2.97$ & $ 5.1$ & $ 6.39\, ,\,\,0.1 $ & $ 0.088 $& $0.23$ & $1.1$&  &  & 606/547                   \\

                       &                 &                                    &  $ 0.31 $ & $  1.7 \times 10^{-5}$ &  &  & $ 0.99  $          \\

Model 7  $^{c}$& $1.865$ & $ 2.97$ & $ 5.0$ & $ 6.39 \, ,\,\,0.1 $ & $  0.085$& $0.22$ & $0.97$& $1.0$&--& 570/544                   \\

                       &                 &                                    &  $0.22 $ & $ 1.7 \times 10^{-5}$ &  & & $ 0.84  $& 2.97    \\ \\

Model 8   $^{c}$& $1.865 \pm 0.02$ & $ 2.97$ & $ 5_{-0.8}^{+0.7}$ & $ 6.39^{+0.03}_{-0.03} ,\,0.1^{+0.03}_{-0.03} $ & $ 0.085^{+ 0.002}_{-0.002}$& $0.22 \pm 0.02$ &$ 0.97_{-0.17}^{+0.13}$ &$1^{+0.4}_{-0.11}$ & $ 0.565 _{-0.009}^{+0.008},\, 0.01_{-0.003}^{+0.002}$ & 559/541                   \\

    (Best fit)     &                 &                                    &  $ 0.22^{+0.02}_{-0.07} $ & $ 1.7^{+0.5}_{-0.7} \times 10^{-5} $ & &  & $ 0.84_{-0.12}^{+0.25} $&$2.97 \pm 0.11 $ &  $1.0^{+0.6}_{-0.5} \times 10^{-4} $       \\ \hline

\end{tabular} 
\end{center}
{\begin{flushleft}

$^{a}${\scriptsize  PL= Power Law, wabs= Galactic absorption, PCA= partially covering cold absorption, BB= Black Body, WA= Warm absorbers developed using Cloudy, GL=Gaussian line.}

$^{b}${\scriptsize These models are in the range 2-10 keV; $^{c}$These models are in the range 0.3-10 keV}

$^{d}$ {\scriptsize  ${\rm Model\, 1= wabs\times PL ;\, Model\, 2=wabs\times PCA\times PL;\,Model \,3=wabs\times PCA\times (PL+GL);\, Model\, 4=wabs\times PCA\times (PL+GL+BB)}$ }\\

{\scriptsize${\rm Model 5=wabs\times PCA\times (PL+GL+BB(1)+BB(2));\,  Model 6=wabs\times PCA\times WA \times (PL+GL+BB(1)+BB(2)); }  $} \\

{\scriptsize${\rm Model\, 7=wabs\times PCA\times WA(1) \times WA(2)\times(PL+GL+BB(1)+BB(2));\, Model \,8=wabs\times PCA\times WA(1) \times WA(2)\times(PL+GL+BB(1)+BB(2)+GL(2))}  $}

$^{e}${\scriptsize\rm Fluxes are in the units of ${\rm Photons\cmsqi\,s^{-1}}$}

{\scriptsize The observed Luminosities ${\,\rm  L_{0.3-10 \kev}= 4.2_{-0.05}^{+0.05}\times 10^{42} \lunit,\,\,L_{0.3-2 \kev}= 2.02_{-0.02}^{+0.04}\times 10^{42} \lunit, \,\,L_{2-10 \kev}= 2.18_{-0.02}^{+0.02}\times 10^{42} \lunit  }$}

\end{flushleft}
}

}
%\end{table*}

%\end{lscape}
\end{sidewaystable*}

\end{document}